\documentclass{article}

\usepackage{arxiv}

\usepackage[utf8]{inputenc} 
\usepackage[T1]{fontenc}    
\usepackage{hyperref}       
\usepackage{url}            
\usepackage{booktabs}       
\usepackage{amsfonts}       
\usepackage{nicefrac}       
\usepackage{microtype}      
\usepackage{lipsum}
\usepackage{graphicx}
\usepackage{amsmath} 

\usepackage[ruled,vlined]{algorithm2e}

\title{Regressor: A C program for Combinatorial Regressions}

\author{
  Eduardo M. Vasconcelos\\
  Academic Department of Control Electric Systems\\
  Federal Institute of Education, Science and Technology\\
  Recife\\
  \texttt{eduardo.vasconcelos@recife.ifpe.edu.br} \\
   \And
  Adriano Gouveia de Souza \\
  Academic Department of Control Electric Systems\\
  Federal Institute of Education, Science and Technology\\
  Recife, Pernambuco, Brazil\\
  \texttt{adrianosouza@recife.ifpe.edu.br} \\
}

\begin{document}
\maketitle

\begin{abstract}
    In statistics, researchers use Regression models for data analysis and prediction in many productive sectors (industry, business, academy, etc.). Regression models are mathematical functions representing an approximation of dependent variable $Y$ from n independent variables $X_i \in X$. The literature presents many regression methods divided into single and multiple regressions. There are several procedures to generate regression models and sets of commercial and academic tools that implement these procedures. This work presents one open-source program called Regressor that makes models from a specific variation of polynomial regression. These models relate the independent variables to generate an approximation of the original output dependent data. In many tests, Regressor was able to build models five times more accurate than commercial tools.
\end{abstract}


\section{Introduction}

Regression models are mathematical functions representing a dependent variable $Y$ from the n independent variable $X_i \in X$. Many regression techniques in literature can be mentioned, such as Linear, Polynomial, Logistic, Ridge, Larson, and many others \cite{FERNANDEZDELGADO201911}. 

Regressions are used in many scientific areas as well as in business, industry, and so on. In Engineering, for example, models are used to predict systems behavior with particular configurations \cite{CHANTANA20191063}; in business, they are used to estimate further market steps and, in economy, for comprehending indicators behavior \cite{AMOOZADKHALILI202030}.

This work focuses on multiple polynomial regressions since they are widely used and have a simple implementation. Many math commercial programs offer an implementation of polynomial regression such as R, LibreOffice Calc and Excel, most commonly implementing the linear model, that can be represented as a function $\widehat{Y} = \beta_0 + \sum_{i=1}^n \beta_i \times X_{i-1}$, where $\widehat{Y}$ is the estimation of set $Y$ and $\beta = \{\beta_0, \beta_1, \dots \beta_n\}$ is the set of regression coefficients. The linear regression advantage is its ability to indicate the trend of data. The disadvantage is its linearity since this class of models cannot output nonlinear response.

This manuscript presents a C program that uses a variation of multiple polynomial regression to build models relating the $ X $ variables, allowing the model a superior fit to data. The model produced by the algorithm has the following structure $\widehat{Y} = \sum \{ \beta_i \times \prod X_k^ \xi \}$.

\section{Combinatorial Regression}

A combinatorial regression model is a polynomial built through the  combination amongst $X$ variables. This class of polynomial models contains more terms than traditional polynomial regression. Considering a fist degree polynomial regression with two variables, the main structure is $\widehat{Y} = \beta_0 + \beta_1 X_1 + \beta_2 X_2$, and using combinatorial regression and the same polynomial degree the following equation is obtained $\widehat{Y} = \beta_0 + \beta_1 X_1 + \beta_2 X_2 + \beta_3 X_1  X_2$. The additional term $X_1 X_2$ to the first order model, allows a nonlinear response, consequently increase its ability to fit an output data. Figure 1 presents an example of regression using the compound interest formula given by $A = C\times (1+i)^t$, where $A$ is the amount of money earned, $C$ is the initial capital invested, $i$ is the interest and $t$ is the period of investment. The graphical representation with 1000 samples of the linear regression is presented on Figure 1(a), whilst Figure 1(b) represent the combinatorial regression.

\begin{figure}[h]
\centering
\includegraphics[scale=0.3]{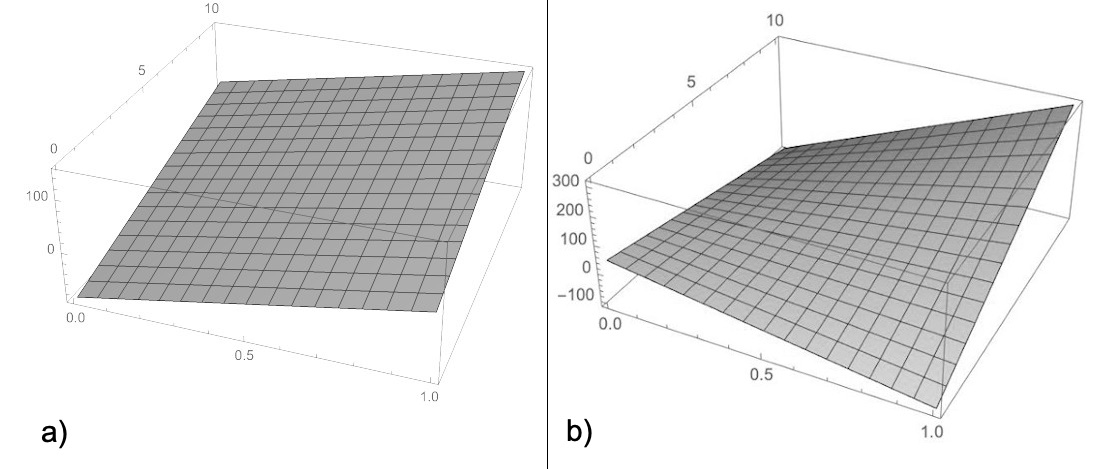}
\caption{\label{fig:1} Linear Regression Examples}{\small{a) linear regression, b) first degree combinatorial model}}
\end{figure}

To build the models of Figure \ref{fig:1} we have defined $C=1$. The models of Figure \ref{fig:1} (a) and (b) are $-91.66 + 116.71i + 12.68t$ and $42.36  - 152.76i - 15.21t + 57.43it$ respectively. The $R^2$ value of the model Figure \ref{fig:1}(a) is $0.34$, meanwhile, the $R^2$ value of Figure \ref{fig:1}(b) model is $0.61$. Note that, in this simple example, the addition of the combination of the term $i \times t$ on the regression, increased the $R^2$ to around $80\%$.
 
\begin{figure}[h]
\centering
\includegraphics[scale=0.3]{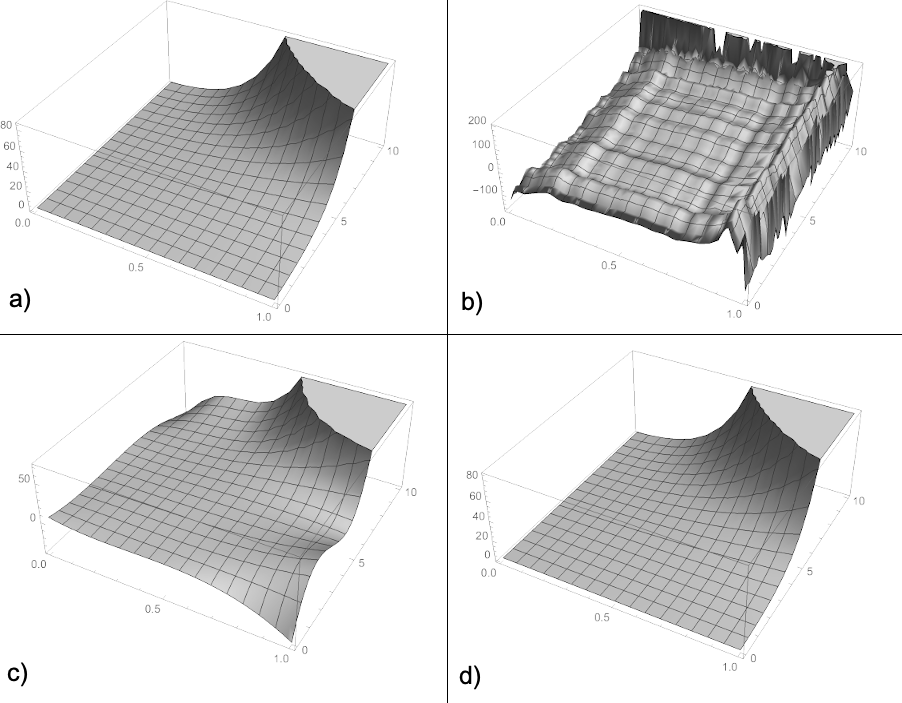}
\caption{\label{fig:2} Polynomial Regression Examples}{\small{a) composed interest formula, b) Fifty degree polynomial model, c) Three degree combinatorial model, (d) Thirty degree combinatorial model}}
\end{figure}

As the linear model is a particular case of a polynomial model, it's possible to improve it accuracy by increasing the polynomial degree. As the linear regression, the use of variables combination in polynomial models increases the fitness of the model. Figure \ref{fig:2} presents a comparison between the polynomial and combinatorial models. Figure \ref{fig:2} (a) present the plot of the compound interest formula, while Figure \ref{fig:2} (b) and (c) represents the polynomial models; (b) polynomial and (c) combinatorial.

Figure \ref{fig:2} (b) is the graphical representation a polynomial regression model with degree $d=50$, its $R^2$ value is 0.46. Even increasing the polynomial degree, the $R^2$ value doesn't reach 0.55, showing that the polynomial regression is better than a linear model, but fails to represent the data set. Figure \ref{fig:2}(c) is a graphical representation of a three degree combinatorial regression; this model has a $R^2 = 0.99$. Comparing the graphs (a) and (c) for Figure \ref{fig:2}, we can observe the regression approximation to the original formula. Finally, Figure \ref{fig:2}(d) presents the graphical representation of a thirty-degree combinatorial regression; for this model $R^2 = 1$, which indicates that the model perfectly fits the samples used to build the model.

\begin{figure}[h]
\centering
\includegraphics[scale=0.5]{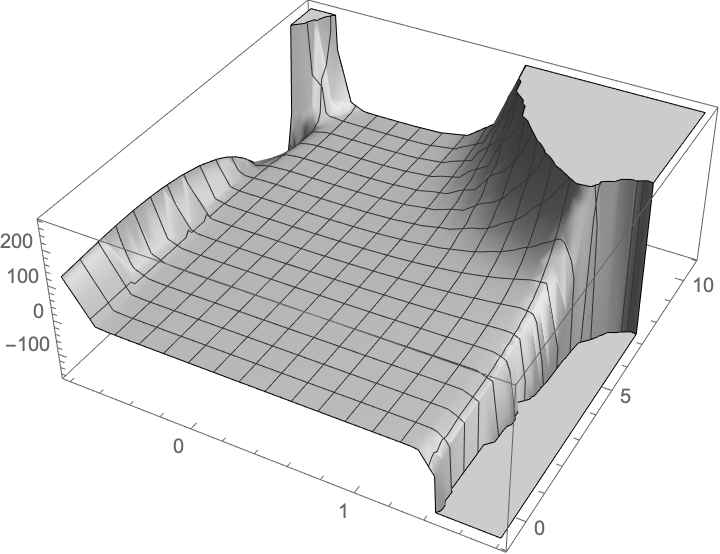}
\caption{\label{fig:3} The divergence of polynomial model}{\small{Thirty degree combinatorial model}}
\end{figure}

It's possible to verify that, as a higher degree, there is higher accuracy using the combinatorial model. But, this concept has two significant disadvantage: (i) the number of terms is determine by $n_t = (d+1)^v$ with $(v)$ variables and $(d)$ polynomial degrees, what indicates that $n_t$ increases exponentially; (ii) the model tends to diverge in its borders as a consequence of higher polynomial degrees, as shown in Figure \ref{fig:3}. The divergence behavior of the polynomial model difficult the prediction, and, as higher the degree is, the model diverges faster.

Despite the combinatorial regression issues, this class of regression fits better than the traditional polynomial regression. Even if the number of terms increases exponentially, the computational complexity of calculating $\widehat{Y}$ is low since the model still comprises of a set of sums and multiplications.

A third issue that has been empirically observed is that the adjustment of these models seems to be low for complex trigonometric functions. But, as in many areas such as economy and business, there are few problems with trigonometric behavior, this problem is not a significant concern.

\subsection{Regression Algorithm}

The algorithm to compose the combinatorial regression is similar to the algorithm for regular polynomial regression, as presented in Equation \ref{eq:1}.

\begin{equation}
\label{eq:1}
B = (M^T M)^{-1}(M^T Y)
\end{equation}

Where $B$ is the matrix of estimated coefficients. The difference to the regular polynomial regression is the construction of matrix $ M $, which uses the following structure:

\begin{equation}
\label{eq:2}
M = 
\begin{vmatrix}
1 & \zeta_{0,1} & \dots & \zeta_{0,n_t} \\
1 & \zeta_{1,1} & \dots & \zeta_{1,n_t}\\
\vdots & \vdots & \ddots & \vdots\\
1 & \zeta_{y,1} & \dots & \zeta_{y,n_t}
\end{vmatrix}
\end{equation}

with

\begin{equation}
\label{eq:3}
    \zeta_{r,c} = \prod_{k=0}^v{X_{r,k}^{\xi(c,k)}}
\end{equation}

and

\begin{equation}
\label{eq:4}
\xi(j,u) = integer \left(\frac{j}{(d+1)^{u}} \right)\times mod (d+1)
\end{equation}

Where $y$ is the number of rows in matrix $Y$, $integer(.)$ is a function that converts real numbers into integers, and $mod(.)$ computes the rest of a division.

The number of columns of the matrix M increases according to the polynomial degree and the number of variables. Besides the before-mentioned issue, models generated through equation \ref{eq:1} presents bad results for large data sets. It happens due to the float-point arithmetic used in computers, that rounds less significant values for large numbers. As to compute equation \ref{eq:1} several multiplications are required, the number of less significant values losses makes the model loses its accuracy. To solve this problem, we implemented the QR factorization to find betas coefficients by solving the system $B=X^{-1} Y$.

The Algorithm \ref{alg:1}, presents the C function $performRegression$. This function generates the matrix $M$, which is represented by the variable $rMatrix$. The variable $limit$ is used to decrease the number of $k$ iterations since, according to equation \ref{eq:3}, in the fists columns of $ M $ the lasts values of $ X $ will not be calculated, allowing to increase the algorithm performance.

\begin{algorithm}[H]
\label{alg:1}
\SetAlgoLined
\caption{Perform Regression Function}

\textbf{double}$*\ performRegression($\textbf{double}$ \ *Y, \ $\textbf{double}$ \ *X,\ $ \textbf{int} $\ degree,\ $ \textbf{int} $\ rows,\ $ \textbf{int}\ $XColumns)\{ $\\

    \quad \textbf{int}$ \ rows = Ylength;$\\
    \quad \textbf{int}$ \ variables = XColumns;$ \\ 
    $  $\\
    \quad \textbf{while} $\ (pow(degree + 1,\ variables) > rows) \{ $
        $degree--;$
    $\}$\\
    $ $\\
    \quad \textbf{int} $\ columns = pow(degree + 1, variables);$\\

    \quad \textbf{double} $\ * rMatrix = ($\textbf{double}$*) malloc (rows*columns*sizeof($\textbf{double}$));$\\
    $ $\\

    \quad \textbf{for} $($\textbf{int} $\ i = 0;\ i < rows;\ i++)\{$\\

       \quad \quad  $*(rMatrix + i) = 1.0;$\\
        $ $\\
        \quad \quad \textbf{for} $($ \textbf{int} $\ j = 1;\  j < columns;\  j++)\{$ \\
            \quad \quad \quad \textbf{double} $product = 1.0;$\\
            
            \quad \quad \quad \textbf{int} $limit = ($\textbf{int}$)(log(j)/log(degree+1))+1;$\\
            $ $\\

            \quad \quad \quad \textbf{for} $($\textbf{int}$\ k = 0;\ k < limit;\ k++)\{ $\\
            $ $\\
               \quad \quad \quad \quad $ product *= pow( *(X + i*variables + k) , ((($\textbf{int}$)((j) / (pow(degree + 1, k)))) \% (degree + 1)));$ \\
            $ $\\
            \quad \quad \quad $\}$\\
            $ $\\
            \quad \quad \quad $*(rMatrix + j * rows + i) = product;$\\
            $ $\\
        \quad \quad $\}$\\

    \quad $\}$\\

    \quad \textbf{return} $qrBetacalculator(rMatrix, Y, rows, columns);$\\
    
$\}$

\end{algorithm}

Function $qrBetacalculator$ is a C modified implementation of Jama Matrix library \cite{JAMA:2020}. We have modified the $ QR $ factorization algorithm to improve its overall performance. Jama Matrix implementation makes row iterations to decompose the matrix, and we adapted the algorithm to make column iterations, allowing the computer to access adjacent memory slots. That is the reason for $rMatrix = M^T$. Executing the Jama Matrix algorithm in a dual-core processor computer with a DDR3 bus memory, the decomposition of a matrix of $45000 \times 625$ dimensions performed in 650 seconds, while the modified algorithm performed in 150 seconds.

\subsection{Regressor}

To demonstrate the performance of combinatorial regression, we built an open-source program called Regressor \footnote{Available in: https://github.com/dr-eduardovasconcelos/Regressor}. This program uses the algorithm presented in algorithm \ref{alg:1} to create models from predefined CSV files. Regressor is executed in the command line, and receives two parameters in its initialization, an integer representing the degree and a string with the path for the CSV file.

The first column of the input CSV file must contain the dependent variable (Y). Use dot (.) to separate the decimals and comma (,) to separate columns. Users have to ensure that there are no empty values in columns so that the program can return an unexpected outcome. 

Regressor generates two files; the first one is a .txt file with the $R^2$ approximation of the model and a spreadsheet format formula for use in software such as LibreOffice Calc\footnote{Tool available in https://www.libreoffice.org}. The second file is a CSV with the residuals of the model.

\section{Data Analysis}

To analyze the performance of models generated by Regressor, we have used four data sets and compared the models with those obtained from LibreOffice Calc multiple regression and models obtained by a modification of Regressor to generate polynomial models. 

The first two data sets were downloaded from UCI Machine Learning Repository\footnote{UCI datasets are available in https://archive.ics.uci.edu/ml/index.php}. The first dataset is entitled "\emph{Physicochemical Properties of Protein Tertiary Structure}" and has 45730 samples with nine variables. The second data set is entitled "\emph{Heart failure clinical records}" and contains 399 rows with 13 variables. As the number of samples in the heart failure data set is small, was generated a second data set with the most significant variables; we used the correlation to rank the variables. The resultant data set was a CSV file with four variables. This process was necessary because the number of rows of $M$ has to be higher than the number of columns.

The third data set was retrieved from a project stored on the GitHub web site entitled \emph{Cancer Prediction Project}\footnote{Cancer data set is available in https://github.com/Arnab777as3uj/STAT6021-Cancer-Prediction-Project}. This data set is part of a challenge for multiple linear regression and has 3048 samples. This data set has, among its information, numerical and symbolical data. Since the purpose of this manuscript is to test the Regressor's performance, we discard the symbolical data. The cancer challenge data set has 26 numerical variables. To this number of variables, it is impossible to generate a combined polynomial with the current number of samples, since $2^{26}$ is higher than 3048. We have produced a secondary data set with seven variables, selecting those variables with a correlation higher than 0.4, following the same process used for generating the secondary heart failure data set.

The fourth data set is available on Kaggle web site\footnote{The COVID-19 data set is available in https://www.kaggle.com/allen-institute-for-ai/CORD-19-research-challenge}, and is a spreadsheet file with 5644 rows containing SARS-CoV-2 testing results and clinical exams from anonymous patients. The Hospital Israelita Albert Einstein has released the data set as part of a challenge aiming to create mechanisms to help identify patients that test positive for COVID-19. The data set has 111 columns, but there are a large number of empty cells. To generate the regression models, we created a secondary data set with 14 variables and 598 rows. As the number of rows is not enough to create a regression with Regressor, we have created a third data set with the nine more relevant parameters.

\subsection{Results}

Table \ref{tab:1} presents a summary of the models created. The polynomial degree used for models generated by Regressor Polynomial\footnote{Available in: https://github.com/dr-eduardovasconcelos/RegressorPolynomial} tool maximizes the $R^2$. In other words, the program can generate models with a higher degree with a lesser $R^2$ value; it is probably due to the float-point roundness on multiplications or the overtraining of the models. For example, for the heart failure with eleven variables and polynomial degree three, the $R^2$ diverged for a number lesser than $-10^{-10}$; it happened due to the high values of the \emph{platelets} variable. 

\begin{table}
 \caption{$R^2$ value of models}
  \centering
  \begin{tabular}{lllll}
    \toprule
    Dataset id & Number of variables & Polynomial degree & Tool & $R^2$ value \\
    \midrule
    Protein Structure & 9 & 1 & LibreOffice calc & 0.2823 \\
          & 9 & 29 & Regressor Polynomial & 0.3624     \\
          & 9 & 1  & Regressor & 0.4305   \\
          & 9 & 2  & Regressor & 0.4503 \\
    Heart Failure  & 11 & 1 & LibreOffice calc & 0.3998 \\
         & 11 & 2 & Regressor Polynomial & 0.4863 \\
         & 4  & 10 & Regressor Polynomial & 0.5708 \\
         & 4  & 1 & Regressor & 0.4099 \\
         & 4 & 3 & Regressor & 0.9481 \\
    Cancer & 26 & 1 & LibreOffice calc & 0.5049 \\
         & 26 & 13 & Regressor Polynomial & 0.6124 \\
         & 7 & 20 & Regressor Polynomial & 0.5430 \\
         & 7 & 1 & Regressor & 0.5548 \\
         & 7 & 2 & Regressor & 0.8925 \\
    COVID-19 & 14 & 1 & LibreOffice calc & 0.1743 \\
         & 14 & 25 & Regressor Polynomial & 0.6010 \\
         & 9 & 1 & Regressor & 0.8815\\
    \bottomrule
  \end{tabular}
  \label{tab:1}
\end{table}

Note that the $R^2$ values for Protein Structure models are low even for models generated with Regressor. The low correlation among the dependent and independent variables can explain this fact. Only two variables have a correlation greater than 0.1, and the higher value is 0.37.

Finally, as it's possible to observe in Table \ref{tab:1}, models generated from Regressor presented better performance than the other models. When comparing the heart failure models, we can see that the Regressor's model with only four variables achieved a $R^2$ two times greater than the model generated by LibreOffice Calc with eleven variables, and this difference has been five times for COVID-19 model.

An important fact about data sets COVID-19 and heart-failure is that their dependent variables are binary: for COVID-19, the positive or negative result for SARS-CoV-2 infection and for heart failure, the death event. Assuming 1 for every predicted absolute value greater than 0.5 and 0 otherwise,  the heart failure model presented four errors in 299 instances; while for the COVID-19 model, there where two errors in 598 cases. These results show that the combinatorial models can adjust better to a set of data, and for binary data sets, the accuracy can reach higher values. 

\section{Conclusions}

In this work, it was presented a program called Regressor that uses the combinatorial regression concept to create multiple variable models. A combinatorial model is a polynomial with multiplications among variables. This class of regression presents best results than the regular polynomial. In the best case, the $R^2$ value generated by Regressor was five times greater than the linear regression feature of LibreOffice Calc, an open-source tool widely used for data analysis.

As future work, we pretend to develop mathematical methods to allowing prediction in high degree combinatorial models.

\bibliographystyle{unsrt}  

\bibliography{references}

\end{document}